\documentclass[hyper]{JHEP} 

\usepackage{epsfig}

\newcommand\fverb{\setbox\pippobox=\hbox\bgroup\verb}

\newcommand\fverbdo{\egroup\medskip\noindent%

            \fbox{\unhbox\pippobox}\ }

\newcommand\fverbit{\egroup\item[\fbox{\unhbox\pippobox}]}

\newbox\pippobox

\title{Note about Static D1-brane in
I-brane Background}

\author{J. Kluso\v{n}
 \footnote{On leave from Masaryk University, Brno}\\
Dipartimento di Fisica,\\
Universita' di Roma \& Sezione di Roma 2, Tor Vergata \\
Via della, Ricerca, Scientifica, 1 00133  Roma   ITALY\\
E-mail:
\email{Josef.Kluson@roma2.infn.it}}

\preprint{ROM2F/2005/24 \\
\hepth{0511304}}

 \abstract{In this short note we will
construct the static solutions on
the world volume of D1-brane embedded in
I-brane background.} 


\def\bz{\mathbf{z}}
\def\by{\mathbf{y}}

\def\bA{\mathbf{A}}

\def\bAi{\left(\mathbf{A}^{-1}\right)}

\def\mL{\mathcal{L}}

\def\tR{S}
\begin{document}
\section{Introduction and Summary}\label{first}
In the recent  paper by  N. Itzaki, D.
Kutasov and N. Seiberg
\cite{Itzhaki:2005tu} the background
that consists two stacks of fivebranes
in type IIB theory that intersect on
$R^{1,1}$ was studied. Among many
interesting results discovered there is
the one that claims that there is an
enhancement of the symmetry in the near
horizon region of the I-brane. More
precisely, it  was shown that in the
near horizon limit  a
 combination of two transverse
 directions  naturally combines with the
world volume dimensions of I-brane
where I-brane is defined classically as
the intersection of two orthogonal
stacks of NS5-branes. For detailed
explanation of this result we recommend
 paper  \cite{Itzhaki:2005tu} for
further reading
\footnote{For related work, see \cite{Lin:2005nh}.}.

In our recent paper
\cite{Kluson:2005eb} we have studied
the time dependent dynamics of D1-brane
probe in this background and we have
given another evidence for an existence
of the enhancement of the symmetry in
the near horizon region. In this paper
we will continue  this study of the
properties of I-brane background from
the point of view of a D1-brane probe.
We focus now on the static solutions on
the world volume theory of  D1-brane.
We perform an analysis in the near
horizon region where the exact static
solution can be found. Following
\cite{Kluson:2005eb}
 we study this
region from two different
 points of view. The first
one is based on an existence of the
additional symmetry on the world volume
of D1-brane that allows us -together
with the existence of the world volume
stress energy tensor- to find exact
static solutions. In the second
approach we firstly perform the
transformations used in
\cite{Itzhaki:2005tu} on the world
volume of D1-brane. This transformation
makes the symmetry enhancement manifest
and the static solution can be easily
found. It turns out that these two
approaches give the same results.

On the other hand as will be
clear from the nature of these solutions
they  cannot be
valid for all points on the
 world volume of D1-brane since at
some points these solutions do not obey
the near horizon approximation. Even if
this an unpleasant property of  these
solutions it is still possible to give
them physical interpretation that is
based on an observation  that this
D1-brane  has many common with the
$AdS_2$-brane studied in
\cite{Bachas:2000fr} and recently in
\cite{Huang:2005hy}. In particular, in
\cite{Huang:2005hy} the static
configuration of $AdS_2$-brane in the
background of $N$ NS5-branes was
studied. We will show, in the same way
as in the paper
\cite{Bachas:2000fr,Huang:2005hy} that
these D-branes cannot reach the
world-volume of I-brane and hence
cannot correspond to the well know
situation when D-brane is stretched in
the transverse directions to the world
volume of NS5-branes
\cite{Elitzur:2000pq,Giveon:1998sr}. On
the other hand we will argue that there
exist solutions of the D-brane
equations of motion that in the near
horizon region corresponds to D1-brane
that ends on the world volume of
I-brane. We find these solutions when
we will study the D1-brane equation of
motion without imposing the gauge where
the spatial coordinate on  the world
volume of D1-brane  is equal to the
spatial coordinate on the world volume
of I-brane. Then we will calculate the
components of the target space stress
energy tensor for these configurations
and we find that the D1-brane wraps
spatial curve in the space transverse
to I-brane.  We will also show that in
the special case when the number of
NS5-branes in two orthogonal stacks is
equal we  can find solutions that
correspond to D1-brane that ends on
I-brane and extends in the whole
transverse space. Unfortunately we were
not able to find such a solution for
general number of NS5-branes in two
orthogonal stacks.

In summary, we again present some
evidence for the existence of the
enhancement symmetry in the near
horizon region of I-brane. The most
natural extension of this approach is
to search for an exact CFT description
of such a D1-brane configuration,
following
\cite{Israel:2005fn,Israel:2003ry}. It
would be also nice to see whether these
 static solutions have some imprint on
the dual little string theory
\cite{Aharony:2004xn}. Our goal is
also to
 understood   the super
symmetric properties of given solution.
In particular it would be nice to see
whether
 they can be
studied using general form of the
calibration conditions
\cite{Gutowski:1999tu,Cascales:2004qp}.

The organization of this paper is as
follows. In the next section
(\ref{second}) we review the basic
facts about D1-brane in I-brane
background and we also present the form
of the conserved charges determined in
\cite{Kluson:2005eb}. Then in section
(\ref{second}) we obtain  the static
solutions of D1-brane equation of
motions that are valid in the near
horizon limit.  Solutions that can be
interpreted as D1-brane that ends on
I-brane will be given in section
(\ref{fourth}).  Finally, in section
(\ref{fifth}) we will determine the
components of the target space stress
energy tensor and evaluate them on the
solutions obtain in previous sections.
\section{Review of basic facts
about D1-brane in the background of
I-brane}\label{second}
 In this section
we briefly review the basic facts about
I-brane background studied recently in
the work \cite{Itzhaki:2005tu}.
This background consists stack of
$k_1$ NS5-branes extended in
$(0,1,2,3,4,5)$ direction and the set
of $k_2$ NS5-branes extended in
$(0,1,6,7,8,9)$ directions. Let us
define
\begin{eqnarray}
\by=(x^2,x^3,x^4,x^5) \ , \nonumber \\
\bz=(x^6,x^7,x^8,x^9) \ .
\nonumber \\
\end{eqnarray}
We will consider the background where
 $k_1$ NS5-branes are localized
at the point $\bz=0$ and $k_2$
NS5-branes localized at the point
$\by=0$. The supergravity background
corresponding to this configuration
takes the form \cite{Khuri:1993ii}
\begin{eqnarray}\label{orb}
\Phi(\bz,\by)=\Phi_1(\bz)+
\Phi_2(\by) \ , \nonumber \\
g_{\mu\nu}=\eta_{\mu\nu} \ ,
\mu,\nu=0,1 \ , \nonumber \\
g_{\alpha\beta}=e^{2(\Phi_2-
\Phi_2(\infty))}\delta_{\alpha\beta} \
, H_{\alpha\beta\gamma}=
-\epsilon_{\alpha\beta\gamma\delta}
\partial^\delta \Phi_2 \ ,
\alpha,\beta,\gamma,\delta=
2,3,4,5\nonumber \\
g_{pq}=e^{2(\Phi_1-\Phi_1(\infty))}
\delta_{pq} \ ,
H_{pqr}=-\epsilon_{pqrs}
\partial^s\Phi_1 \ ,
p,q,r,s=6,7,8,9 \ , \nonumber \\
\end{eqnarray}
where $\Phi$ on the
first line means the dilaton and
where
\begin{eqnarray}
e^{2(\Phi_1-\Phi_1(\infty))}=H_1(\bz)
=1+\frac{k_1l_s^2} {|\bz|^2} \ ,
\nonumber \\
e^{2(\Phi_2-\Phi_2(\infty))}=H_2(\by)
=1+\frac{k_2l_s^2} {|\by|^2} \ .
\nonumber \\
\end{eqnarray}
We start with DBI action for  D1-brane
 that moves in I-brane background
\begin{equation}\label{sga}
S=-\tau_1\int d^2x
e^{-\Phi}\sqrt{-\det\bA} \ ,
\end{equation}
where $\tau_1$ is D1-brane tension. We
restrict ourselves in this section to
the gauge fixed form of the D1-brane
action  where the static gauge is
defined as
\begin{equation}\label{stt}
X^0=x^0 \ ,
X^1=x^1\ .
\end{equation}
More general case will be studied
in  section (\ref{fourth}).
For the static gauge
(\ref{stt})  the matrix
$\bA_{\mu\nu}$ in (\ref{sga})
takes the form
\begin{equation}
\bA_{\mu\nu}=g_{\mu\nu}+
g_{IJ}\partial_\mu X^I
\partial_\nu X^J+
b_{IJ}\partial_\mu X^I
\partial_\nu X^J+\partial_{\mu}A_{\nu}
-\partial_\nu A_\mu \ ,
I,J=2,\dots,9 \ .
\end{equation}
where $X^I$ parameterize the position
of D1-brane in transverse space and
where $A_\mu \ , \mu \ ,\nu=0,1$ are
components of the world volume gauge
field.

It turns out that in order to
effectively analyze the properties of
D1-brane in the I-brane background the
knowledge of various world volume
conserved quantities is  useful.
Natural example is the world volume
stress energy tensor. It can be derived
in various ways, for example one can
introduce auxiliary world volume metric
and perform the variation of the action
with respect to it. It can be also
determined  by standard Noether
procedure. This approach was used in
our previous paper \cite{Kluson:2005eb}
and we got the result
\begin{eqnarray}\label{stressenergytensor}
 T^\mu_\nu=\tau_1e^{-\Phi}
\left(\delta^\mu_\nu-g_{IJ}\partial_\kappa
 X^J\bAi^{\kappa\mu}_S\partial_\nu X^I-
b_{IJ}\partial_\kappa X^J\bAi^{\kappa\mu}_A
\partial_\nu X^I-\right. \nonumber \\
\left.
-\partial_\nu A_\sigma\bAi^{\sigma\mu}_A\right)
\sqrt{-\det\bA} \ ,
\nonumber \\
\end{eqnarray}
where
\begin{equation}
\bAi^{\nu\mu}_S=
\frac{1}{2}\left(\bAi^{\nu\mu}+
\bAi^{\mu\nu}\right)\ ,
\bAi^{\nu\mu}_A=
\frac{1}{2}\left(\bAi^{\nu\mu}-
\bAi^{\mu\nu}\right)\ ,
\end{equation}
and where the components $T^\mu_\nu$
obey the   equation
\begin{equation}\label{sete}
\partial_\mu T^\mu_\nu=0 \ .
\end{equation}
In case when we have one dynamical
variable the knowledge of $T^\mu_\nu$
and the equation (\ref{sete}) are
sufficient for determining the possible
trajectory or static solution of given
D1-brane. However it turns out that the
dynamics of D1-brane in I-brane
background is characterized by two
dynamical modes. Even if  we could in
principle find the trajectories of
given D1-brane by solving the equation
of motion it is very difficult task.

The second possibility how to find possible
D1-brane configurations is  to search for
another current that obeys the
partial differential equation of
the first order.
As was  shown in
\cite{Kluson:2005eb} such a current
can be found in the
near horizon limit of I-brane background.
Since the near horizon limit
is defined as
\begin{equation}
\frac{k_1l_s^2}
{|\bz|^2}\gg 1 \ ,
\frac{k_2l_s^2}
{|\by|^2}\gg 1 \
\end{equation}
  we can write
\begin{equation}
H_1=\frac{\lambda_1}
{|\bz|^2} \ , \lambda_1=k_1l_s^2 \ ,
H_2=\frac{\lambda_2}
{|\by|^2} \ , \lambda_2=k_2l_s^2 \
\end{equation}
and consequently
the  action (\ref{sga})
takes the form
\begin{equation}\label{D1nhl}
S=-\tau_1\int d^2x
\frac{1}{g_1g_2\sqrt{H_1H_2}}
\sqrt{-\det \bA} \ ,
\end{equation}
where
\begin{eqnarray}\label{A1nh1}
\bA_{\mu\nu}=
\eta_{\mu\nu}+H_1\delta_{pr}
\partial_\mu Z^p\partial_\nu Z^r+
H_2\delta_{\alpha\beta}\partial_\mu
Y^\alpha\partial_\nu Y^\beta+\nonumber \\
+B_{pr}\partial_\mu Z^p\partial_\nu Z^r
+B_{\alpha\beta}\partial_\mu Y^\alpha
\partial_\nu Y^\beta+\partial_{\mu}A_{\nu}
-\partial_\nu A_\mu \ ,
\nonumber \\
\end{eqnarray}
and where $g_1=e^{\Phi_1(\infty)} \ ,
g_2=e^{\Phi_2(\infty)}$. The form
of the action (\ref{D1nhl}) and
the matrix (\ref{A1nh1}) suggests
that  it
is natural to consider following
transformations
\begin{equation}\label{Gtr}
Z'^p(x')=\Gamma Z^p(x) \  ,
Y'^\alpha(x')=\Gamma^{-1}Y^\alpha(x) \
, A'_\mu(x')=A_\mu(x) \ , x'^\mu=x^\mu
\ ,
\end{equation}
where $\Gamma$ is a real parameter. It
can be shown that this
transformation implies, under some
conditions, an existence of conserved
charge. In fact, there exists a current
$j^\mu_D$  related to the
transformation  (\ref{Gtr}) that has
the form
\begin{eqnarray}\label{dilcg}
j^\mu_D=
\tau_1e^{-\Phi}
\left(g_{pr}Z^p\partial_\nu Z^r
\bAi^{\nu\mu}_S+b_{pr}Z^p
\partial_\nu Z^r\bAi^{\nu\mu}_A
\right)\sqrt{-\det\bA}-\nonumber \\
-\tau_1e^{-\Phi}
\left(g_{\alpha\beta}Y^\alpha
\partial_\nu Y^\beta\bAi^{\nu\mu}_S
+b_{\alpha\beta}Y^\alpha
\partial_\nu Y^\beta \bAi^{\nu\mu}_A
\right)\sqrt{-\det\bA} \
\nonumber \\
\end{eqnarray}
and that obeys the
equation
\begin{equation}\label{jdde}
\partial_\mu j^\mu_D=
 \tau_1e^{-\Phi}
\left(b_{pq}\partial_\mu Z^p
\partial_\nu Z^q-b_{\alpha\beta}
\partial_\mu Y^\alpha\partial_\nu
Y^\beta\right)\bAi^{\nu\mu}\sqrt{-\det
\bA} \ .
\end{equation}
Note that the components $g$ and $b$ given
above correspond to their near horizon
limit. As follows from the right hand side of the
equation (\ref{jdde}), the conservation
of $j^\mu_D$ is restored for homogenous
modes or for modes that are spatial
depend only since then the right side
of the equation above vanishes thanks
to the antisymmetry of $b$. The case of
the pure time dependent modes  was
analyzed in our previous paper
\cite{Kluson:2005eb}. The static
solutions will be studied in the next
section.
\section{Static Solutions in the
near horizon limit}\label{third} In
this section we will discuss the
spatial dependent solution of D1-brane
in I-brane background where all modes
(except $A_1$) do not depend on $x^0$.
Thanks to the manifest rotation
symmetry $SO(4)$ in the subspaces
spanned by coordinates
$\bz=(z^6,z^7,z^8,z^9)$ and
$\by=(y^2,y^3,y^4,y^5)$ we can reduce
the problem to the study of the motion
in two dimensional subspaces, namely we
will presume that only following world
volume modes are excited
\begin{equation}
z^6=R \cos \theta \ , z^7=R \sin \theta
\ ,
\end{equation}
and
\begin{equation}
y^2=\tR \cos\psi \ , y^3=\tR \sin\psi \
.
\end{equation}
 There
is also the  gauge field $A_1$ present
and we  work in the  gauge $A_0=0$.

Now using the fact that the action does
not depend on angular modes explicitly
it follows from the equation of motion
that
\begin{equation}
\partial_1\left[e^{-\Phi}
g_{\theta\theta}\partial_1\theta
\bAi^{11}_S\sqrt{-\det\bA}\right]=0
\end{equation}
that can be solved with
$\theta=\theta_0=\mathrm{const}$. The
same result holds for $\psi$ as well
and for that reason we will not discuss
these modes at all.

Then for  spatial dependent modes $S,R$
and for nonzero $A_1$ we get following
components of the matrix $\bA$
\begin{equation}
\bA_{00}=-1 \ ,
\bA_{10}=F_{10}=-\dot{A}_1=-\bA_{01}  \
, \bA_{11}=1+H_1R'^2+H_2S'^2 \ ,
\end{equation}
where $(\dots)'\equiv \partial_1(\dots)$.
Then
\begin{equation}\label{detbas}
\det\bA=-(1+H_1R'^2+H_2S'^2)+
(\partial_0 A_1)^2
\end{equation}
and also
\begin{equation}
\bAi=\frac{1}{\det\bA}
\left(\begin{array}{cc}
1+H_1R'^2+H_2S'^2 & \partial_0 A_1 \\
-\partial_0 A_1 & -1 \\ \end{array} \right)
\ .
\end{equation}
Using these results  the components of
the stress energy tensor
(\ref{stressenergytensor}) take the
form
\begin{eqnarray}
T_0^0= -\frac{\tau_1e^{-\Phi}
(1+H_1R'^2+H_2S'^2)}{\sqrt{-\det\bA}} \
,
\nonumber \\
T_0^1=0 \ ,
T_1^0=-\frac{\tau_1e^{-\Phi}
\partial_1 A_1\partial_0 A_1}{\sqrt{-\det\bA}}
\nonumber \\
T_1^1=\tau_1e^{-\Phi}
\frac{(-1+(\partial_0
A_1)^2)}{\sqrt{-\det\bA}} \ .
\nonumber \\
\end{eqnarray}
Firstly the  equation (\ref{sete}) for
$\nu=0$ implies
\begin{equation}
\partial_0 T^0_0=0
\end{equation}
that of course is trivially satisfied
as $T^0_0$
 does not depend
on $x^0$ by construction. On the other
hand the equation (\ref{sete}) for
$\nu=1$ gives
\begin{equation}
\partial_0 T^0_1+\partial_1T^1_1 \ ,
\end{equation}
where
\begin{equation}
\partial_0 T^0_1=
-\frac{\tau_1e^{-\Phi}\partial_1
\partial_0 A_1\partial_0 A_1}
{\sqrt{-\det\bA}} \ .
\end{equation}
We see that the presence of nonzero
$\partial_0 A_1$ makes the analysis
more complicated. We will return to
this problem later and now we restrict
to the case when  $\partial_0 A_1=0$.
Then we also get $T^0_1=0$ and
consequently
\begin{equation}\label{t11s0}
\partial_1 T^1_1=0
\Rightarrow
T^1_1(x^1)=-\frac{\tau_1}{g_1g_2
\sqrt{H_1H_2}\sqrt{1+H_1R'^2+H_2S'^2}}=K
\ ,
\end{equation}
where $K$ is arbitrary constant.
 To
find solution with nontrivial $S$ and
$R$ we again restrict to the near
horizon region where
\begin{equation}
H_1=\frac{\lambda_1}{R^2} \ ,
H_2=\frac{\lambda_2}{S^2} \ .
\end{equation}
As we have argued in the previous
section we can find in this region the
dilatation current (\ref{dilcg}) that
obeys the differential equation
(\ref{jdde}).  Using the fact that the
only dynamical modes are $S,R$ we see
that the right side of this equation
vanishes.  It is also easy to see that
$j^0_D=0$ and the conservation of the
$j^\mu_D$ reduces into
\begin{equation}\label{pjd}
\partial_1 j^1_D=0 \ ,
\end{equation}
where
\begin{eqnarray}
j^1_D=-\frac{\tau_1}{g_1g_2\sqrt{H_1H_2}}
\left(H_1RR'-H_2SS'\right)\sqrt{-\det\bA}=
\nonumber \\
=-\frac{\tau_1RS}{g_1g_2\sqrt{\lambda_1
\lambda_2}}
\left(\frac{\lambda_1}{R^2}RR'-
\frac{\lambda_2}{S^2}SS'\right)
\frac{1}{\sqrt{1+H_1R'^2+H_2S'^2}} \ .
\nonumber \\
\end{eqnarray}
It follows from the equation
(\ref{pjd})  that $j^1_D$ does not
depend on $x^1$. Let us then denote its
value as $j^1_D\equiv D$. Using
(\ref{t11s0}) we can express $D$ as
\begin{equation}\label{Ds}
D=K\left(\frac{\lambda_1 R'}{R}-
\frac{\lambda_2S'}{S}\right) \ .
\end{equation}
By integration of  this equation we get
\begin{equation}
\frac{D}{K}x^1+D_0=
\ln R^{\lambda_1}-\ln S^{\lambda_2} \
\end{equation}
or equivalently
\begin{equation}\label{SRs1}
S=R^{\lambda_1/\lambda_2}e^{-\frac{1}{\lambda_2}
(\frac{D}{K}x^1+D_0)} \ .
\end{equation}
To find the spatial dependent solution
we use (\ref{Ds}) to express $S'$ as
function of $R,R'$ and $S$ and then we
insert it to the expression for $T^1_1$
(\ref{t11s0})
\begin{equation}
-\frac{\tau_1
RS}{\sqrt{\lambda_1\lambda_2}
g_1g_2}\frac{1}{\sqrt{
1+\lambda_1\frac{R'^2}{R^2}+
\frac{1}{\lambda_2}\left(\lambda_1\frac{R'}{R}-
\frac{D}{K}\right)^2}}=K \
\end{equation}
that leads to the quadratic equation in
the variable $R'$
\begin{equation}
\frac{\lambda_1^2}{\lambda}
\frac{R'^2}{R^2}-2\frac{\lambda_1}{\lambda_2}
\frac{D}{K}\frac{R'}{R}
+1+\frac{D^2}{\lambda_2 K^2}
-\frac{\tau_1^2}{K^2\lambda_1 \lambda_2
(g_1g_2)^2}R^2S^2=0 \ ,
\end{equation}
where
$\frac{1}{\lambda}=\frac{1}{\lambda_1}+\frac{1}{\lambda_2}$.
 This equation   has two  roots
\begin{eqnarray}\label{R'}
R'=\frac{D\lambda R}{K\lambda_1
\lambda_2} \pm\frac{\lambda R}{
\lambda_1^2}
\sqrt{\frac{\tau_1^2\lambda_1}
{K^2(g_1g_2)^2\lambda_2\lambda}R^2S^2
 -\frac{\lambda_1^2}{\lambda}-\frac{\lambda_1}{\lambda_2}
\frac{D^2}{K^2}}\ .  \nonumber \\
\end{eqnarray}
We will solve the equation (\ref{R'})
with the ansatz
\begin{equation}\label{Tc}
R(x^1)=C(x^1)
e^{\frac{D\lambda}{K\lambda_1\lambda_2}x^1}
\ .
\end{equation}
If we insert (\ref{Tc}) into the
equation (\ref{R'}) we obtain a
differential equation for $C$ in the
form
\begin{equation}\label{C'}
C'=\pm\frac{\lambda}{\lambda_1^2}
\sqrt{ \frac{\lambda_1
\tau_1^2e^{-2\frac{D_0}{\lambda_2}}
}{K^2(g_1g_2)^2\lambda_2\lambda}
C^{\frac{2\lambda_1}{\lambda}}
-\left(\frac{\lambda_1^2}{\lambda}
+\frac{\lambda_1}{\lambda_2}\frac{D^2}{K^2}\right)}
\ ,
\end{equation}
where we have used the relation
\begin{equation}
R^2S^2=R^{2(\lambda_1+\lambda_2)/\lambda_2}
e^{-\frac{2}{\lambda_2}
(\frac{D}{K}x^1+D_0)}=C^{\frac{2\lambda_1}{\lambda}}
e^{-2\frac{D_0}{\lambda_2}}
\end{equation}
that follows
 from (\ref{SRs1}). After
straightforward integration of the
equation (\ref{C'}) we obtain following
spatial dependence of $R$ on $x^1$
\begin{equation}\label{RsD}
R=e^{\frac{1}{\lambda_1+\lambda_2}
(\frac{D}{K}x^1+D_0)}
\left(\frac{\sqrt{\lambda_1
\lambda_2}K(g_1g_2)\sqrt{
1+\frac{1}{\lambda_1+\lambda_2}\frac{D^2}{K^2}}}
{\tau_1}\right)^{\frac{\lambda}
{\lambda_1}}
\left(\frac{1}{\cos
\frac{1}{\sqrt{\lambda}}\sqrt{1+
\frac{D^2}{(\lambda_1+\lambda_2)K^2}}x^1}
\right)^{\frac{\lambda}{\lambda_1}} \ .
\end{equation}
Before we try to give a physical
meaning of this result  we determine
this solution  from a different point
of view using the transformations
proposed in \cite{Itzhaki:2005tu}. Let
us again start with the Lagrangian
density for D1-brane that is inserted
in the near horizon region and again
restrict to the case of the spatial
dependent radial modes only
\begin{equation}\label{actcom}
\mL=-\frac{\tau_1}{g_1g_2}\sqrt{\frac{R}{\lambda_1}}
\sqrt{\frac{S}{ \lambda_2}}
\sqrt{1+\frac{\lambda_1}{R^2}
(\partial_1
R)^2+\frac{\lambda_2}{S^2}(\partial_1
S)^2} \ .
\end{equation}
As the first step let us introduce two
modes $\phi_1$ and $\phi_2$ as
\begin{equation}\label{RSp}
R=e^{\frac{\phi_1}{\sqrt{\lambda_1}} }\
, S=e^{\frac{\phi_2}{\sqrt{\lambda_2}}}
\end{equation}
so that the Lagrangian (\ref{actcom})
takes the form
\begin{equation}\label{accom1}
\mL=-\frac{\tau_1e^{
(\frac{\phi_1}{\sqrt{\lambda_1}} +
\frac{\phi_2}{\sqrt{\lambda_2}})}
}{\sqrt{\lambda_1\lambda_2}g_1g_2}
\sqrt{1+(\partial_1\phi)^2+
(\partial_1\phi)^2} \ .
\end{equation}
Then we  introduce two modes $ \phi,
x^2$ defined as
\begin{equation}\label{defpx}
Q\phi=\frac{1}{\sqrt{\lambda_1}}\phi_1+
\frac{1}{\sqrt{\lambda_2}}\phi_2 \ ,
Qx^2=\frac{1}{\sqrt{\lambda_2}}\phi_1-
\frac{1}{\sqrt{\lambda_1}}\phi_2 \ ,
\end{equation}
where
\begin{equation}
Q=\frac{1}{\sqrt{\lambda}} \ ,
\frac{1}{\lambda}=\frac{1}{\lambda_1}+
\frac{1}{\lambda_2} \ .
\end{equation}
Now  using the inverse transformations
to (\ref{defpx})
\begin{eqnarray}\label{phi1i}
\phi_1=\frac{1}{\sqrt{\lambda_1+\lambda_2}}
\left(\sqrt{\lambda_1}x^2+\sqrt{\lambda_2}\phi
\right) \ , \nonumber \\
\phi_2=\frac{1}{\sqrt{\lambda_1+\lambda_2}}
\left(\sqrt{\lambda_1}\phi-
\sqrt{\lambda_2}x^2\right) \  \nonumber \\
\end{eqnarray}
the  Lagrangian  (\ref{accom1}) takes
the form
\begin{equation}\label{accoms}
\mL=-\frac{\tau_1}{\sqrt{\lambda_1\lambda_2}
g_1g_2}e^{Q\phi}\sqrt{1+(\partial_1 x^2)^2
+(\partial_1 \phi)^2} \ .
\end{equation}
 Since this Lagrangian does not
explicitly depend on $x^1$ it follows
that the quantity $T^1_1=\frac{\delta
\mL}{\delta \partial_1 x^2}
\partial_1 x^2+\frac{\delta \mL}{\delta
\partial_1 \phi}\partial_1\phi-\mL$
obeys the equation $\partial_1 T^1_1=0$
that again implies
\begin{equation}\label{Kphi}
-\frac{\tau_1}{\sqrt{\lambda_1\lambda_2}
g_1g_2}e^{Q\phi}\frac{1}{
\sqrt{1+(\partial_1 x^2)^2 +(\partial_1
\phi)^2}}=K \ .
\end{equation}
Also using the fact that the Lagrangian
does not depend on $x^2$ we immediately
obtain following spatial independent
quantity
\begin{equation}
\partial_1\left(\frac{\delta \mL}
{\delta \partial_1 x^2}\right)=0
\Rightarrow \frac{\delta \mL}
{\delta \partial_1 x^2}=P_2=\mathrm{const}
\end{equation}
that allows us to write
\begin{equation}\label{px2}
 \partial_1
x^2=\frac{P_2}{K}
\end{equation}
and hence
\begin{equation}
x^2=\frac{P_2}{K}x^1+x^2_0 \ .
\end{equation}
If we insert (\ref{px2}) into
 to the equation (\ref{Kphi})
  we obtain the differential
equation for $\phi$ that has the form
\begin{equation}
\frac{d\phi}
{\sqrt{\frac{\tau_1^2e^{2Q\phi}}
{\lambda_1\lambda_2(g_1g_2)^2K^2}
-\left(1+\frac{P^2_2}{K^2}\right)}}=
\pm dx^1
\end{equation}
and that has the solution
\begin{equation}\label{Qs}
e^{Q\phi}=\frac{\tau_1}{\sqrt{\lambda_1
\lambda_2}g_1g_2
K\sqrt{1+\frac{P^2_2}{K^2}}}
\frac{1}{\cos Q
\sqrt{1+\frac{P^2_2}{K^2}} x^1} \ .
\end{equation}
If  we perform an identification
 between $P_2$ and $D$ in the form
\begin{equation}
P_2=\frac{D}{
\sqrt{\lambda_1+\lambda_2}} \
\end{equation}
and use  the  relations (\ref{phi1i})
we can map  the solution (\ref{Qs}) to
the solution (\ref{RsD}). In other
words, we have two equivalent
descriptions of D1-brane in the near
horizon region of I-brane. In the first
one, that is valid for D1-brane in the
near horizon region of the original
background (\ref{orb}) the world volume
action posses additional, scaling like
symmetry. In the second one, based on
the background introduced in
\cite{Itzhaki:2005tu} we get that the
world volume theory is manifestly
invariant in the direction $x^2$ and
consequently the momentum $P_2$ is
conserved. In both cases an enhancement
of the symmetry in the near horizon
region is observed.

Finally we return to the case when
$\partial_0 A_1\neq 0$. Firstly, recall
that the equations of motion for
$A_\mu$ take the form
\begin{equation}\label{Aeqs}
\partial_\nu \left[e^{-\Phi}\bAi^{\nu\mu}_A\sqrt{-\det\bA}
\right]=0 \ ,
\end{equation}
or explicitly
\begin{eqnarray}\label{eqAe}
\partial_1\left[\frac{1}{g_1g_2
\sqrt{H_1H_2}}\frac{\partial_0 A_1}
{\sqrt{1+H_1 R^2+H_2S'^2-(\partial_0
A_1)^2}}\right]=0 \ , \nonumber \\
\partial_0\left[\frac{1}{g_1g_2
\sqrt{H_1H_2}}\frac{\partial_0 A_1}
{\sqrt{1+H_1 R^2+H_2S'^2-(\partial_0
A_1)^2}}\right]=0 \ . \nonumber \\
\end{eqnarray}
The second equation can be solved with
$\partial_0^2A_1=0$ while the first
equation can be written as
\begin{equation}
\partial_1[\partial_0 A_1 T^1_1]=\partial_1 \partial_0
A_1T_1^1+
\partial_0 A_1\partial_1 T^1_1=0 \
\end{equation}
that has again solution  $\partial_1
A_1=0$ and $\partial_1 T^1_1=0$. This
result together with the first equation
in (\ref{eqAe}) implies that the
quantity $\partial_0 A_1 T^1_1\equiv
-\pi$ is constant and we can write
\begin{equation}
\partial_0 A_1=
-\frac{\pi}{K} \ .
\end{equation}
Then we obtain following form of the
constant $T^1_1$
\begin{equation}
-\frac{\tau_1 RS}{ g_1g_2
\sqrt{\lambda_1\lambda_2}}
\frac{1}{\sqrt{1+\frac{\lambda_1}
{R^2}R'^2+ \frac{\lambda_2}{S^2}
S'^2-\frac{\pi^2}{K^2}}}=K \ ,
\end{equation}
where we have taken the near horizon
limit in the end. Now we see that this
equation has exactly the same form as
the equation given in case when $\pi=0$
with the small difference that the
constant term under the square root is
not $1$ but $1-\frac{\pi^2}{K^2}$. Then
we could find the static solution
exactly in the same way as above.

Now we return to the question of the
physical interpretation of the spatial
dependent solution given in (\ref{Qs})
or in  (\ref{RsD}). We mean that it
corresponds to
 an array of spikes  that approach to the
world volume of I-brane at distance
$R_{min},S_{min}$ at the points $
\left(1+\frac{D^2}{(\lambda_1+
\lambda_2)K^2}\right)x_k=k\pi ,
k=0,1,\dots$ and then they extend in
$R$ and $S$ directions until they reach
the distance where the near horizon
approximation ceases to be valid. In
some sense these solutions correspond
to $AdS_2$-branes studied in paper
\cite{Bachas:2000fr} where the
classical solution was valid for the
whole $AdS_3$. This analysis was
extended to the study of $AdS_2$ branes
in the paper \cite{Huang:2005hy} and
the results given there are similar to
our solutions.

It is also important to stress that we
cannot find smooth solution that
describes  D1-brane that terminates on
the world volume of I-brane. In fact,
the static gauge defined in the
previous section is not suited  for
description of this configuration. In
order to find such D1-brane static
solution we should consider the general
form
 of D1-brane effective action
without where the static gauge is not
imposed.
\section{DBI action without static
gauge presumption}\label{fourth}
In this section we present an
alternative form of the embedding of
the D1-brane  in I-brane background.
The starting point of our analysis is
Dirac-Born-Infeld action for Dp-brane
in a general background
\begin{eqnarray}\label{actae}
S=-\tau_p\int d^{p+1}\sigma
e^{-\Phi}\sqrt{-\det \bA} \ ,
\bA_{\mu\nu}=\gamma_{\mu\nu}+F_{\mu\nu}
\ ,
\nonumber \\
\end{eqnarray}
where $\tau_p$ is Dp-brane tension,
$\Phi(X)$ is dilaton and where
$\gamma_{\mu\nu} \ , \mu,\nu=0,\dots,
p$ is embedding of the metric to the
world volume of Dp-brane
\begin{equation}
\gamma_{\mu\nu}=g_{MN}\partial_\mu X^M
\partial_\nu X^N \ , M,N=0,\dots, 9 \ .
\end{equation}
In (\ref{actae}) the form $F_{\mu\nu}$
is defined as
\begin{equation}
F_{\mu\nu}=b_{MN}\partial_\mu X^M
\partial_\nu X^N+\partial_\mu A_\nu -
\partial_\nu A_\mu \ .
\end{equation}
Note also that we label the world
volume coordinates of Dp-brane with
$\sigma^\mu \ , \mu=0,\dots,p$.

The equations of motion for $X^K$
can be easily determined from
(\ref{actae}) and take the form
\begin{eqnarray}
\partial_K[e^{-\Phi}]\sqrt{-\det\bA}
+\frac{1}{2}e^{-\Phi}\left(
\partial_K g_{MN}
\partial_\mu X^M\partial_\nu X^N
+\partial_K b_{MN}\partial_\mu
X^M\partial_\nu X^N\right)
\bAi^{\nu\mu}\sqrt{-\det\bA}-
\nonumber \\
-\partial_\mu\left[e^{-\Phi}
g_{KM}\partial_\nu X^M\bAi^{\nu\mu}_S
\sqrt{-\det\bA}\right]
-\partial_{\mu}\left[e^{-\Phi}
b_{KM}\partial_\nu X^M\bAi^{\nu\mu}_A\sqrt{-\det\bA}
\right]=0 \ .
\nonumber \\
\end{eqnarray}
Finally, we should also determine the equation of motion
for the gauge field $A_\mu$:
\begin{equation}\label{Aeq}
\partial_\nu \left[e^{-\Phi}\bAi^{\nu\mu}_A\sqrt{-\det\bA}
\right]=0 \ .
\end{equation}
Using these equations of motion we present an
alternative solutions corresponding to
the  static D1-brane
in I-brane background. First of all we presume
that all modes (except $A_1$, where we however
demand that $\partial_0^2A_1=0$) depend on
$\sigma^1$ only. Since we consider
time independent solution it is
natural to fix
\begin{equation}
X^0=\sigma^0 \ .
\end{equation}
 Then using an antisymmetry of
$b_{MN}$ it is clear that its embedding
to the world volume of D1-brane
is zero and consequently the matrix
$\bA_{\mu\nu}$ takes the form
\begin{equation}
\bA=\left(\begin{array}{cc}
-1  &\partial_0 A_1 \\
-\partial_0 A_1 & g_{IJ}\partial_1
X^I\partial_1 X^J \\ \end{array}
\right) \
\end{equation}
hence the determinant and  the inverse
matrix are equal to
\begin{equation}
\det\bA=-g_{IJ}\partial_1 X^I
\partial_1 X^J+(\partial_0 A_1)^2 \ ,
\end{equation}
\begin{equation}
\bAi=\frac{1}{\det\bA}
\left(\begin{array}{cc}
g_{IJ}\partial_1 X^I\partial_1 X^J
 & -\partial_0 A_1 \\
\partial_0 A_1 & -1 \\ \end{array}
\right)\ ,
\end{equation}
where $I,J=1,\dots,9$.
Then the equation of motion
(\ref{Aeq}) implies
\begin{eqnarray}
\partial_0 \left[e^{-\Phi}\bAi^{10}_A
\sqrt{-\det\bA}
\right]=-\partial_0
\left[\frac{1}{\sqrt{H_1H_2}g_1g_2}
\frac{\partial_0 A_1}{\sqrt{-\det\bA}}
\right]=0  \nonumber \\
\partial_1 \left[e^{-\Phi}\bAi^{01}_A
\sqrt{-\det\bA}
\right]=\partial_1
\left[\frac{1}{\sqrt{H_1H_2}g_1g_2}
\frac{\partial_0 A_1}{\sqrt{-\det\bA}}
\right]=0 \ , \nonumber \\
\end{eqnarray}
The first equation is satisfied if we
presume that $\partial_0 A_1$ does not
depend on time. The second one implies
that $\pi\equiv
\frac{1}{\sqrt{H_1H_2}g_1g_2}
\frac{\partial_0 A_1}{\sqrt{-\det\bA}}$
does not depend on $\sigma^1$. In
summary, we will solve these equations
with $\pi=const$. For simplicity of
resulting formulas we take $\pi=0$.

As in previous section we use the
manifest rotation symmetry $SO(4)$ in
the subspaces spanned by coordinates
$\bz=(z^6,z^7,z^8,z^9)$ and
$\by=(y^2,y^3,y^4,y^5)$ so that we  presume
that  only following world
volume modes are excited
\begin{equation}
z^6=R \cos \theta \ , z^7=R \sin \theta
\ ,
\end{equation}
and
\begin{equation}
y^2=\tR \cos\psi \ , y^3=\tR \sin\psi \
.
\end{equation}
Since the metric does not depend
on $\theta,\psi$ the equation of motions
for them reduce into
\begin{eqnarray}
\partial_1\left[e^{-\Phi}g_{\theta
\theta}\partial_1 \theta
\bAi^{11}\sqrt{-\det\bA}\right]=0
\ , \nonumber \\
\partial_1\left[e^{-\Phi}g_{\psi
\psi}\partial_1 \psi
\bAi^{11}\sqrt{-\det\bA}\right]=0
\ , \nonumber \\
\end{eqnarray}
that can be solved with the ansatz
\begin{equation}
\psi=\psi_0=\mathrm{const} \ ,
\theta=\theta_0=\mathrm{const} \ .
\end{equation}
Recall  that there  exists  the mode
$X^1$ on the world volume of D1-brane
that parameterizes its position  along
the
 world volume of I-brane.
 This mode obeys the equation of
motion
\begin{equation}
\partial_1\left[e^{-\Phi}g_{11}
\partial_1 X^1
\bAi^{11}_S\sqrt{-\det\bA}\right]=0 \ .
\end{equation}
If we denote $X^1(\sigma^1)=f$ then the
equation above implies
\begin{eqnarray}
\frac{1}{g_1g_2\sqrt{H_1}\sqrt{H_2}}
\frac{f'}{\sqrt{H_1 (\partial_1
R)^2+H_2(\partial S)^2+(\partial_1
f)^2}}=K \Rightarrow
\nonumber \\
(\partial_1
f)^2=\frac{H_1(\partial_1R)^2+H_2(\partial_1S)^2}
{\frac{1}{(g_1g_2)^2K^2H_1H_2}-1} \ .
\nonumber \\
\end{eqnarray}
From the last equation we get following
condition
\begin{equation}
\frac{1}{K^2(g_1g_2)^2}>H_1H_2 \ .
\end{equation}
This condition implies that for
$K\neq 0$  $R,S$ cannot be equal to zero.
In other words for $K\neq 0$ we cannot
find solution that
corresponds to the D1-brane that
ends  on the world volume of a
I-brane. We have seen the similar
result in the previous section.
Then in order to find
D1-brane that ends on I-brane we should
take $K=0$ that however
 implies $\partial_1
X^1=0$ and hence $X^1=x^1_0$.

As we could expected the problem
reduced to the study
of the equation of motion  for
$S,R$.  However in general case
we are not able to proceed further
without an existence of an
additional symmetry. On the other hand
we can gain some information about the
solution that describes D1-brane that
ends on I-brane if we restrict to the
near horizon limit where we can
either  perform the
transformation introduced in
\cite{Itzhaki:2005tu} or use
the scaling like symmetry found
in \cite{Kluson:2005eb}.
Since these two approaches are
equivalent we restrict ourselves
in this section to the first
possibility. We again
start with the Lagrangian
density of D1-brane in the
near horizon region of I-brane
\begin{equation}\label{actcomns}
\mL=-\frac{\tau_1}{g_1g_2}
\sqrt{\frac{R^2}{\lambda_1}}\sqrt{\frac{S^2}{
\lambda_2}}\sqrt{\frac{\lambda_1}{R^2}
(\partial_1 R)^2+\frac{\lambda_2}{S^2}
(\partial_1 S)^2} \ .
\end{equation}
Observe that this Lagrangian density
differs from the Lagrangian density
given in (\ref{actcom}) in the fact
that the factor $1$ is absent in
(\ref{actcomns}). Now  we  introduce
two modes $\phi_1$ and $\phi_2$ defined
 as
\begin{equation}\label{RSp}
R=e^{\frac{\phi_1}{\sqrt{\lambda_1}} }\
, S=e^{\frac{\phi_2}{\sqrt{\lambda_2}}}
\end{equation}
and hence the Lagrangian
(\ref{actcomns}) takes the form
\begin{equation}\label{accom1ns}
\mL=-\frac{\tau_1e^{
(\frac{\phi_1}{\sqrt{\lambda_1}} +
\frac{\phi_2}{\sqrt{\lambda_2}})}
}{\sqrt{\lambda_1\lambda_2}g_1g_2}
\sqrt{(\partial_1\phi)^2+
(\partial_1\phi)^2} \ .
\end{equation}
Again using the transformations
(\ref{phi1i}) we can rewrite the
Lagrangian density (\ref{accom1ns})
into the more symmetric form
\begin{equation}\label{accomsns}
\mL=-\frac{\tau_1}{\sqrt{\lambda_1\lambda_2}
g_1g_2}e^{Q\phi}\sqrt{(\partial_1
x^2)^2 +(\partial_1 \phi)^2} \ .
\end{equation}
Since we did not fix the world volume
coordinate $\sigma^1$ to be equal to
some target space one it turns out that
the spatial component of the world
volume stress energy tensor $T^1_1$
vanishes identically. For that reason
we should directly
 solve the equations of motion
for $x^2$ and $\phi$ that follow
 (\ref{accomsns}). Firstly, using
the fact that the Lagrangian does not
depend on $x^2$ we immediately obtain
following spatial independent quantity
\begin{equation}
\partial_1\left(\frac{\delta \mL}
{\delta \partial_1 x^2}\right)=0
\Rightarrow \frac{\delta \mL} {\delta
\partial_1 x^2}=\frac{\tau_1 e^{Q\phi}}
{\sqrt{\lambda_1\lambda_2}g_1g_2}
\frac{\partial_1 x^2}
{\sqrt{(\partial_1 x^2)^2 +(\partial_1
\phi)^2}}= P_2 \ .
\end{equation}
From this equation we can also
express $\partial_1 x^2$ as
\begin{equation}
(\partial_1 x^2)^2=
\frac{P^2_2(\partial_1 \phi)^2}
{\frac{\tau_1^2}{\lambda_1
\lambda_2(g_1g_2)^2}e^{2Q\phi}-P_2^2} \ .
\end{equation}
We  are interested in the D1-brane that
reaches the world volume of I-brane
that occurs in the limit
$\phi\rightarrow -\infty$. On the other
hand the equation above implies the
lower bound on the value of $\phi$ in
the form
\begin{equation}
\frac{\tau_1^2}
{\lambda_1\lambda_2(g_1g_2)^2}e^{2Q\phi}
-P_2^2>0
\end{equation}
that can be obeyed in the limit $
\phi\rightarrow -\infty$ in case when
$P_2=0$ that however also
implies  $\partial_1 x^2=0$.

Then  the  equation of
motion for $\phi$ takes the form
\begin{equation}\label{eqphi}
-\frac{\tau_1 Q}{\sqrt{\lambda_1
\lambda_2}g_1g_2}e^{Q\phi}\sqrt{
(\partial_1 x^2)^2 +(\partial_1
\phi)^2}+
\partial_1\left[
\frac{\tau_1e^{Q\phi}}{\sqrt{\lambda_1\lambda_2}(g_1g_2)}
\frac{\partial_1
\phi}{\sqrt{(\partial_1 x^2)
+(\partial_1 \phi)^2}}\right]=0 \ .
\end{equation}
It is easy to see that
the equation of motion
(\ref{eqphi}) can
be solved with the
ansatz (for $\partial_1 x^2=0$)
\begin{equation}\label{phif}
\phi(\sigma^1)=f(\sigma^1) \ ,
\end{equation}
where $f$ is any smooth function.
 Recall that in the
$R,S$ variables this
solution takes the form
\begin{equation}\label{RSf}
R=e^{\sqrt{\frac{\lambda_2}
{\lambda_1(\lambda_1+\lambda_2)}}f(\sigma^1)} \ ,
S=e^{\sqrt{\frac{\lambda_1}
{\lambda_2(\lambda_1+\lambda_2)}}f(\sigma^1)} \ .
\end{equation}
In order to gain better physical
meaning of given solution we should
calculate quantities that do not depend
on the form of the function of $f$.
Examples of such quantities are space
time stress energy tensor or the
currents corresponding to the coupling
of D1-brane to Ramond-Ramond two form
of to NS two form. We restrict
ourselves to the calculation of the
space time stress energy tensor that
will be performed in section
(\ref{fifth}). Before we do this we
would like to show that there exists
the solution that is valid for all
$R,S$. This solution can be found in
case when $\lambda_1=\lambda_2$ as we
will see in the next subsection.
\subsection{The case $\lambda_1=\lambda_2
=\lambda$} We start with the case
$\lambda_1=\lambda_2=\lambda $. We
would like to show that in this case
the DBI equation of motion have the
solution
\begin{equation}\label{RSls}
R=\sqrt{\lambda}g(\sigma^1) \ ,
S=\sqrt{\lambda}g(\sigma^1) \ ,
\end{equation}
for any smooth function $g(\sigma^1)$.
As a justification for the choose of
this ansatz note that the action is
invariant under exchange $R,S$. For
this ansatz we easily get
\begin{equation}\label{H12}
H_1=H_2=H \ , H_1(\partial_1R)^2+
H_2(\partial_2S)=2\lambda Hg'^2 \ ,
\end{equation}
where $H=1+\frac{1}{g^2}$. For
simplicity we again restrict to the
case of $\pi=0$. It can be shown
however that the same solution exists
for $\pi\neq 0$ as well. Now if we
insert the ansatz (\ref{RSls}) and use
(\ref{H12}) to the equation of motion
for $R$ we get
\begin{eqnarray}\label{Rsrga}
\partial_R[e^{-\Phi}]\sqrt{-\det\bA}+
\frac{1}{2}e^{-\Phi}\partial_Rg_{RR}
(\partial_1R)^2\bAi^{11}\sqrt{-\det\bA}
-\nonumber \\
-\partial_1\left[e^{-\Phi}
g_{RR}\partial_1R\bAi^{11}\sqrt{-\det\bA}
\right]=
\frac{1}{\sqrt{2}H^{3/2}}\frac{\lambda^{3/2}g'}{R^3}
-\frac{\lambda^{3/2}
g'}{\sqrt{2}R^3H^{3/2}} =0 \nonumber \\
\end{eqnarray}
that shows that (\ref{RSls}) solves the
equation of motion for $R$.
 Due to the symmetry
between $S$ and $R$ it is clear that
the equation of motion for $S$ is
obeyed as well.

In summary, in case of when the number
of NS5-branes in two orthogonal stacks
is the same we were able to find the
static configuration corresponding to
D1-brane extended in the space
transverse to I-brane world volume.
Since however this solution is given in
terms of any function $g( \sigma^1)$
its interpretation is not completely
clear. In order to gain better insight
to its physical properties we will
calculate the space time stress energy
tensor for this configuration.
\section{Space time stress energy
tensor}\label{fifth} In this section we
determine  components of the space time
stress energy tensor and evaluate them
on the solutions determined in the
previous section.  To begin with we
again write the action as
\begin{equation}
S=-\tau_1\int d^{10}x
d^2\sigma e^{-\Phi}\sqrt{-\det\bA}
\delta^{(10)}(X^M(\sigma)-x^M) \ .
\end{equation}
Since the space time stress energy tensor
is defined as
\begin{equation}
T_{MN}(x)=-\frac{2}{\sqrt{-g(x)}}
\frac{\delta S}{\delta g^{MN}(x)}
 \end{equation}
we get
\begin{equation}\label{Tsp}
T_{MN}=-\frac{\tau_1}{\sqrt{-g(x)}}
\int d^2\sigma e^{-\Phi}
g_{MK}\partial_\mu X^K
g_{NL}\partial_\nu X^L\bAi^{\nu\mu}
\sqrt{-\det\bA}\delta
^{(10)}(X^M(\sigma)-x^M) \ .
\end{equation}
In our case we have $X^0=\sigma^0$.
Then the integral over $\sigma^0$ can
be easily performed and it is equal to
one. Then for the spatial dependent
ansatz and for $\partial_0A_1=0$ we get
\begin{eqnarray}\label{Tgen}
T_{00}=\frac{\tau_1} {\sqrt{-\det
g(x)}} \int d\sigma^1 \frac{\sqrt{
H_1(\partial_1
R')^2+H_2(\partial_1S)^2}}{g_1g_2
\sqrt{H_1}\sqrt{H_2}}
\delta(R(\sigma^1)-u) \delta
(S(\sigma^1)-v) \ ,
\nonumber \\
T_{uu}=-\frac{\tau_1} {\sqrt{-\det
g(x)}} \int d\sigma^1
\frac{H_1^2(\partial_1R)^2}{g_1g_2
\sqrt{H_1}\sqrt{H_2}\sqrt{
H_1(\partial_1
R')^2+H_2(\partial_1S)^2}}
\delta(R(\sigma^1)-u)
\delta (S(\sigma^1)-v) \ ,  \nonumber \\
T_{vv}=-\frac{\tau_1} {\sqrt{-\det
g(x)}} \int d\sigma^1
\frac{H_2^2(\partial_1S)^2}{g_1g_2
\sqrt{H_1}\sqrt{H_2}\sqrt{
H_1(\partial_1
R')^2+H_2(\partial_1S)^2}}
\delta(R(\sigma^1)-u)
\delta (S(\sigma^1)-v) \ ,  \nonumber \\
T_{uv}=-\frac{\tau_1} {\sqrt{-\det
g(x)}} \int d\sigma^1
\frac{H_1H_2\partial_1R\partial_1S}{g_1g_2
\sqrt{H_1}\sqrt{H_2}\sqrt{
H_1(\partial_1
R')^2+H_2(\partial_1S)^2}}
\delta(R(\sigma^1)-u)
\delta (S(\sigma^1)-v) \ ,  \nonumber \\
\end{eqnarray}
where  we have omitted the delta
functions that express that the
components of the stress energy tensor
are localized at fixed values
$x^1_0,\theta_0,\psi_0$. For the
solution $R=g(\sigma^1)\ ,
S=g(\sigma^1)$ we get following result
\begin{eqnarray}
T_{00}(u,v)=\frac{1}{\sqrt{-\det g(u,v)}}
\frac{\tau_1\sqrt{2}}{ g_1g_2}
\frac{1}{\sqrt{1+\frac{\lambda}{u^2}}}\delta(u-v) \ ,
\nonumber \\
T_{uu}(u,v)=-\frac{1}{\sqrt{-\det g(u,v)}}
\frac{\tau_1}{ g_1g_2\sqrt{2}}
\sqrt{1+\frac{\lambda}{u^2}}
\delta(u-v) \ ,
\nonumber \\
T_{vv}(u,v)=-\frac{1}{\sqrt{-\det g(u,v)}}
\frac{\tau_1}{ g_1g_2\sqrt{2}}
\sqrt{1+\frac{\lambda}{u^2}}
\delta(u-v) \ ,
\nonumber \\
T_{uv}(u,v)=-\frac{1}{\sqrt{-\det g(u,v)}}
\frac{\tau_1}{ g_1g_2\sqrt{2}}\sqrt{1+\frac{\lambda}{u^2}}
\delta(u-v) \ .
\nonumber \\
\end{eqnarray}
In these calculations  we have used the
fact that the factor $g'(\sigma^1)$ is
present in all components of the stress
energy tensor and hence we could
perform  the substitution
$g(\sigma^1)=m$. Then the integration
over $m$ swallows up one delta function
$\delta(m-u)$ so that we have replaced
everywhere $m$ with $u$. This implies
that the second delta function becomes
$\delta(u-v)$. We see from the  results
given above  that all components of the
stress energy tensor are localized
around the line $u=v$. In other words
the  configuration (\ref{RSls})
describes D1-brane that is stretched in
directions $u$ and $v$ that are
transverse to the world volume of
I-brane.

Now we will calculate the components of
the stress energy tensor for the
solution  (\ref{RSf}) that is valid in
the near horizon region of I-brane. As
follows from   (\ref{RSf}) we easily
get
\begin{equation}
H_1(\partial_1R)^2+H_2(\partial_1S)^2=
\frac{\lambda_1}{R^2}(\partial_1R)^2
+\frac{\lambda_2}{S^2}(\partial_1S)^2
=f'^2\ ,
\end{equation}
where we have used
\begin{equation}
(\partial_1R)^2=\frac{\lambda_2}
{\lambda_1(\lambda_1+\lambda_2)}f'^2R^2 \ ,
(\partial_1S)^2=\frac{\lambda_1}
{\lambda_2(\lambda_1+\lambda_2)}f'^2S^2 \ .
\end{equation}
Then the components of the space time
stress energy tensors evaluated in the
near horizon approximation are equal to
\begin{eqnarray}
T_{00}=\frac{\tau_1}
{\sqrt{-\det g(x)}}
\int d\sigma^1
\frac{RSf'}{g_1g_2
\sqrt{\lambda_1\lambda_2}}
\delta(R(\sigma^1)-u)
\delta (S(\sigma^1)-v) \ ,
\nonumber \\
T_{vv}=-\frac{\tau_1}
{\sqrt{-\det g(x)}}
\int d\sigma^1
\frac{R\sqrt{\lambda_1\lambda_2}f'}
{S(\lambda_1+\lambda_2)}
\delta(R(\sigma^1)-u)
\delta (S(\sigma^1)-v) \nonumber \\
T_{uu}=-\frac{\tau_1}
{\sqrt{-\det g(x)}}
\int d\sigma^1
\frac{S\sqrt{\lambda_1\lambda_2}f'}
{Rg_1g_2(\lambda_1+\lambda_2)}
\delta(R(\sigma^1)-u)
\delta (S(\sigma^1)-v) \nonumber \\
T_{uv}=-\frac{\tau_1}
{\sqrt{-\det g(x)}}
\int d\sigma^1
\frac{\lambda_1\lambda_2f'}
{g_1g_2(\lambda_1+\lambda_2)}
\delta(R(\sigma^1)-u)
\delta (S(\sigma^1)-v) \ .  \nonumber \\
\end{eqnarray}
To evaluate $T_{00}$ we perform the
 substitution
$R(\sigma^1)=m $ and then we integrate
over $m$ with the result
\begin{eqnarray}
T_{00}=\frac{\tau_1}{\sqrt{-\det g}}
\frac{\sqrt{\lambda_1+\lambda_2}}
{\lambda_2}S(R^{-1}(u))\delta(
S(R^{-1}(u))-v)=\nonumber \\
=\frac{\tau_1}{\sqrt{-\det g}}
\frac{\sqrt{\lambda_1+\lambda_2}}
{\lambda_2}u^{\frac{\lambda_1}{\lambda_2}}
\delta(u^{\frac{\lambda_1}{\lambda_2}}-v)
\ ,\nonumber \\
\end{eqnarray}
using
\begin{equation}
S(R^{-1}(u))=e^{\frac{\lambda_1}{\lambda_2}\ln u}=
u^{\frac{\lambda_1}{\lambda_2}} \ .
\end{equation}
In the same way we can
proceed with  $T_{vv}$
and we get
\begin{equation}
T_{vv}=-\frac{\tau_1}{\sqrt{-\det g}}
\frac{\lambda_1}{\sqrt{\lambda_1+
\lambda_2}u^{\frac{\lambda_1}{\lambda_2}}}
\delta(u^{\frac{\lambda_1}{\lambda_2}}-v) \ .
\end{equation}
On the other hand for $T_{vv}$ we
use the substitution $S(\sigma^1)=m$
so that we get
\begin{eqnarray}
T_{uu}=-\frac{\tau_1}
{\sqrt{-\det g}}\frac{\lambda_2}
{\sqrt{\lambda_1+\lambda_2}R(S^{-1}(v))}
\delta(R(S^{-1}(v))-u)=
\nonumber \\
=-\frac{\tau_1}
{\sqrt{-\det g}}\frac{\lambda_2}
{\sqrt{\lambda_1+\lambda_2}v^{\frac{\lambda_2}
{\lambda_1}}}
\delta(v^{\frac{\lambda_2}{\lambda_1}}-u) \ .
\nonumber \\
\end{eqnarray}
Finally, for $T_{uv}$ we obtain the
result
\begin{eqnarray}
T_{uv}=-\frac{\tau_1}
{\sqrt{-\det g}}
\frac{\lambda_1\sqrt{\lambda_1\lambda_2}}
{\sqrt{\lambda_1+\lambda_2}R(R^{-1}(u))}
\delta(S(R^{-1}(u))-v)=\nonumber \\
=-\frac{\tau_1} {\sqrt{-\det g}}
\frac{\lambda_1\sqrt{\lambda_1\lambda_2}}
{\sqrt{\lambda_1+\lambda_2}u}
\delta(u^{\frac{\lambda_1}{\lambda_2}}-v)
\ .
\nonumber \\
\end{eqnarray}
As follows from all components of the
space time stress energy tensor the
D1-brane in the near horizon limit
spans the curve
$u^{\lambda_1}=v^{\lambda_2}$. We also
see that this result does not depend on
the form of the function
 $f$ given in  (\ref{RSf}) which can be
 interpreted as a consequence of the
 diffeomorphism invariance of D1-brane
 effective action.
\\
\\
{\bf Acknowledgement}

This work
 was supported in part by the Czech Ministry of
Education under Contract No. MSM
0021622409, by INFN, by the MIUR-COFIN
contract 2003-023852, by the EU
contracts MRTN-CT-2004-503369 and
MRTN-CT-2004-512194, by the INTAS
contract 03-516346 and by the NATO
grant PST.CLG.978785.


\end{document}